\documentclass[aps,prb,twocolumn,groupedaddress,floats,showpacs]{revtex4}
\usepackage{latexsym}
\usepackage{dcolumn}
\usepackage[dvips]{graphicx}
\usepackage{amssymb}
\usepackage{graphics}
\usepackage{amsmath}
\usepackage{epsf}

\begin{document}
\author{E. H. Hwang, S. Adam and S. Das Sarma}
\title{Transport in chemically doped graphene in the presence of adsorbed 
molecules}
\affiliation{Condensed Matter Theory Center, Department of Physics, 
University of Maryland, College Park, MD 20742-4111, USA}
\date{\today}
\begin{abstract}
Motivated by a recent experiment reporting on the possible application
of graphene as sensors, we calculate transport properties of 2D
graphene monolayers in the presence of adsorbed molecules.  We find
that the adsorbed molecules, acting as compensators that partially
neutralize the random charged impurity centers in the substrate,
enhance the graphene mobility without much change in the carrier
density.  We predict that subsequent field-effect
measurements should preserve this higher mobility for both electrons
and holes, but with a voltage induced electron-hole asymmetry that
depends on whether the adsorbed molecule was an electron or hole donor
in the compensation process.  We also calculate the low density
magnetoresistance and find good quantitative agreement with
experimental results.
\end{abstract}
\maketitle

\section{Introduction}
\label{Sect:Intro}

The recent discovery of graphene -- a single
layer of graphite -- followed by the rapid progress in fabricating
transistor-like devices and measuring its transport
properties~\cite{kn:novoselov2005,kn:zhang2005} 
is an
important, perhaps seminal, development in low dimensional electronic
phenomena in nanostructures.  These systems are conceptually novel, where
the low-energy description for a single sheet of Carbon atoms in a 
honeycomb lattice is the linear ``Dirac-like'' dispersion having 
both electron and hole carriers (which are the positive and negative
chiral solutions of the Dirac Hamiltonian).  The band structure
induced carrier spectrum in graphene monolayers is four-fold
degenerate (spin and valley), and the intrinsic sublattice symmetry
(two inequivalent Carbon atoms in the unit cell) causes a suppression
of backscattering.  Both of these features could have application in
technology where the former provides the opportunity to have both
spin-tronic and valley-tronic functionality on the same device, and
the theoretical absence of backscattering has led to the speculation
that, as a matter of principle, carrier mobilities in 2D graphene
monolayers could be extremely high even at room temperature.

\begin{figure}
\bigskip
\epsfxsize=0.9\hsize
\hspace{0.0\hsize}
\epsffile{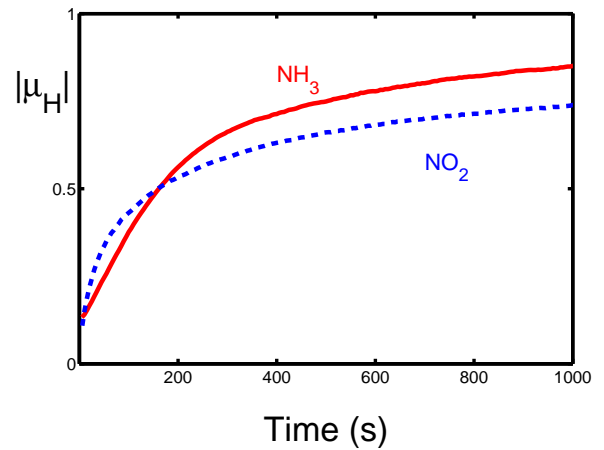}
\caption{\label{Fig:mob} (Color online) Hall mobility $|\mu_H|$ in
units of ${\rm m}^2/{\rm Vs}$ measured in the experiments of
Ref.~\onlinecite{kn:schedin2006a} by chemically doping graphene with
NH$_3$ and NO$_2$.  Notice that $|\mu_H|$ increases as one increases
the dopant exposure time $t$, indicating a decrease in impurity
scattering.}
\end{figure}

\begin{figure}[h]
\bigskip
\epsfxsize=0.9\hsize
\hspace{0.0\hsize}
\epsffile{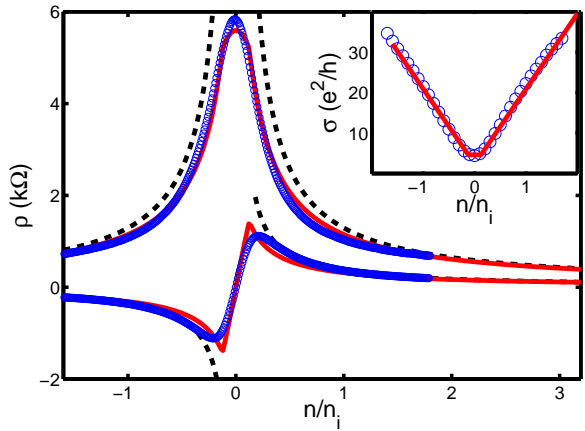}
\caption{\label{Fig:rho} (Color online) Graphene magnetoresistivity 
 $\rho_{xx}$ and $\rho_{xy}$ for $B=1 T$.  Dashed (solid)
lines are calculated for one (two) carrier model, 
Eq.~\ref{Eq:twochannel} with 
impurity concentration $n_i=1.75 \rm{x}10^{12} \rm{cm}^{-2}$. 
Open circles show data of Ref.~\onlinecite{kn:schedin2006a,kn:schedin2006}.
Inset compares longitudinal conductivity 
$\sigma_{xx}$ for the two carrier model (solid line) 
and experiments of Ref.~\onlinecite{kn:schedin2006a,kn:schedin2006} 
(circles).}
\end{figure}

The increase in graphene carrier mobility induced by absorbed gas
molecules has been recently used as a highly-sensitive solid-state
sensor capable of detecting individual
molecules.~\cite{kn:schedin2006a} Notwithstanding the interpretation
made in Ref.~\onlinecite{kn:schedin2006a,kn:schedin2006}, the experimental data
reported there certainly shows Hall mobility $\mu_H =
\rho_{xy}/(\rho_{xx} B)$ increasing with dopant exposure time.  The
experimental data is reproduced in Fig.~\ref{Fig:mob} where for both
$NH_3$ and $NO_2$ adsorbants one finds a strong increase in mobility
$\mu_H$ for small doping concentrations and that this increase in
mobility persists as one increases the dopant exposure time $t$.  It
is further reported in Refs.~\onlinecite{kn:schedin2006,kn:wehling2007} that
the hall mobility then saturates for even larger doping concentrations
or during subsequent annealing of the samples.

While this experimental advance has the potential to revolutionize gas
sensors, it also raises fundamental questions about the mechanism by
which adsorbates change the transport properties of graphene.  Similar
experiments done on Carbon nanotubes~\cite{kn:kong2000,
kn:bradley2003} interpreted their data to argue that $NH_3$ adsorbents
(in the presence of water) transfered $0.04 e^-$ per molecule, while
$N0_2$ binds to the surface and withdraws $0.1 e^-$ per
molecule.~\cite{kn:snow2006}  However, these experiments on Carbon
Nanotubes had no direct way to measure the carrier density, and were
unable to provide further clues as to the charge transfer mechanism.
In contrast, the recent experiments of
Ref.~\onlinecite{kn:schedin2006a,kn:schedin2006} 
on ``chemically-doped'' graphene
(which is a zero-gap semiconductor with both electron and hole
carriers) could provide definitive answers to these long standing
theoretical questions.  Although there has been considerable recent
theoretical activity~\cite{kn:hwang2006c,kn:nomura2007,kn:koshino2006,kn:ando2006,kn:cheianov2006,
kn:verges2007,kn:aleiner2006}
studying carrier transport in graphene, transport in chemically doped
graphene -- the subject of our current work -- has not been considered
previously in the literature.  Since it is now 
believed~\cite{kn:tan2007,kn:chen2007b} that charged impurities
are the dominant scattering mechanism in graphene, one of the main goals 
of the present work is to understand chemical doping of graphene within
the framework of charged impurity scattering.  

We emphasize that the carrier mobility $\mu$ is the average velocity 
of each electron per unit applied electron field.  It is strictly defined 
only in the limit of vanishing electric and magnetic fields.  However,
since for small fields, the  electrical conductivity is the ratio of 
the current $J$ to the electric field $E$, we have the
relation between conductivity and mobility as $\sigma = ne \mu$. 
Therefore, the correct way to obtain the mobility from the 
measured conductivity is to divide the conductivity by a carrier 
density, i.e. $\mu = \sigma/ne$.  Only in the very special case
where conductivity varies strictly linearly with carrier density
is this result the same as the derivative of conductivity with respect 
to density, i.e. $(1/e)d\sigma/dn$, but in general the mobility can be 
quite different from $(1/e)d\sigma/dn$ when the conductivity is not a 
linear function of density.  To our knowledge, the theoretical 
justification for a linear-in-density conductivity is valid only 
for incoherent transport of uncorrelated Coulomb impurities located 
exactly at the surface of the graphene sheet.  Other scattering 
mechanisms likely to be present in graphene including phonons, 
ripples, and defects are known to change the density dependence of the
conductivity necessitating the consistent use of $\mu = \sigma/ne$
when discussing carrier mobility and especially when discussing
multiple scattering mechanisms.  This is relevant in comparing our
theory with the data of Ref.~\onlinecite{kn:schedin2006} where the
mobility has been defined as $(1/e)d\sigma/dn$ rather than the conventional
$\sigma/ne$.

We note that for most graphene samples, the measured conductivities are 
linear over wide range of density.  This fact alone is the smoking gun for 
the dominance of charged impurity scattering in current graphene samples, 
although we note that other signatures of charged impurity scattering 
have been observed in recent experiments.~\cite{kn:tan2007, kn:chen2007b}  
We note that
even within the charged impurity scattering model, the
linear-in-density behavior breaks down at low densities near the Dirac
point, where instead one expects a constant conductivity
plateau  whose value is determined by the residual
density induced by the charged impurities.~\cite{kn:adam2007a}  
At high densities
(i.e. high gate voltages) one expects~\cite{kn:hwang2006c} the competing
effects of short-range scattering (such as
point-defects or dislocations) and shifting of the charged impurities
that also cause the conductivity to deviate from the the linear relation.
For all these reasons, to get the correct mobility one has to divide
the conductivity by the density. 

In a finite magnetic field, the Hall mobility is often used to
characterize samples and is obtained by measuring the conductivity
$\sigma$ and then dividing it by the concentration $n$ found by the
Hall effect. The Hall mobility has the same dimension as the drift
mobility discussed above, but differs by a factor $r_H$, which is the
scattering factor of the Hall coefficient, defined by $r_H = \langle
\tau^2 \rangle/ \langle \tau \rangle ^2$, where $\tau$ is a
relaxation time and $\langle \cdots \rangle$ indicates the energy average.
For a model with only charged impurity scattering and away from the
Dirac point and at zero temperature, one can calculate that the 
difference between the definitions of Hall mobility and drift mobility 
is less than 3
percent, making one a good approximation for the other.  But again,
this is only true away from the Dirac point, and in the case of
charged impurity scattering being the dominant scattering mechanism.
In general, to get the correct Hall mobility experimentally, one would
have to divide the measured Hall conductivity by the carrier density
instead of doing a derivative.

We now proceed as follows.  In Section~\ref{Sect:magneto} we first
generalize the theory~\cite{kn:hwang2006c} of charged impurity
scattering to include the effect of a magnetic field and compare our
theory for graphene magnetoresistance with the experimental results of
Ref.~\onlinecite{kn:schedin2006a} in the absence of chemical dopants.
Having understood the zero doping regime, in Section~\ref{Sect:mobinc}
we discuss the most natural assumption for the effect of chemical
adsorbents which is to transfer charge to the graphene layer leaving
behind a charged impurity.  We find that this always decreases the
graphene mobility in contrast to the experimental findings in
Ref.~\onlinecite{kn:schedin2006a}.  In Section~\ref{Sect:mobdec} we
propose a compensation model to explain the increase in mobility seen
experimentally.  In addition, we highlight that this model makes
several predictions that can be verified experimentally.  In
Section~\ref{Sect:mobconst} we phenomenologically account for the
observed~\cite{kn:wehling2007} saturation of mobility during annealing
experiments on molecular doped graphene sheets.  We note that in our
model a saturation of mobility necessarily means that there is no
change in the charged impurity concentration and we demonstrate that
including an asymmetric shift in threshold voltage reproduces all the
features observed experimentally.  We then conclude in
Section~\ref{Sect:conclusion} with a discussion.              
   
\section{Graphene Magnetoresistance}
\label{Sect:magneto}

Before considering chemical doping issues, we first discuss graphene 
magnetoresistance.  The transport
properties of graphene in the presence of a magnetic field was
considered for ``white noise'' (i.e. delta-correlated) disorder by
Zheng and Ando~\cite{kn:zheng2002}, and in the Landau quantization
limit by Gusynin and Sharapov.~\cite{kn:gusynin2005}  As has been 
demonstrated theoretically in 
Refs.~\onlinecite{kn:nomura2007,kn:cheianov2006,kn:hwang2006c},
and experimentally in Refs.~\onlinecite{kn:tan2007, kn:chen2007b},
the dominant transport mechanism in graphene is Coulomb scattering
from charged impurities (not white noise disorder), and below we 
extend this formalism to incorporate
the effects of a finite (but weak) magnetic field using the Boltzmann 
transport theory in the semiclassical regime where Landau quantization
is unimportant.  While we are concerned 
with the high density limit, where $n > n_i$
and where Boltzmann theory is valid, to compare with experiments we
construct a simplified two-component (i.e. electron and hole) 
model consistent with the
percolation model suggested in Ref.~\onlinecite{kn:hwang2006c}.  To
phenomenologically account for the conductivity saturation at low
density within the Boltzmann theory framework, we assume that for $V_g
> 0$, the electron and hole densities are 
\begin{eqnarray}
\label{Eq:twochannel}
n_e &=& n_0 + n, \nonumber \\
n_h &=& (n_0 -n)\theta(n_0-n),
\end{eqnarray}
respectively, where $\theta(x)$ is the Heavyside step function and $n_0 \sim
n_i$ is the minimum carrier density.  We have similar expressions for
$V_g < 0$.  In this simplified model, $n_0 = 0$ for the one carrier
model and we use $2 n_0 = n_i/4$ for the two 
carrier model.~\cite{kn:adam2007a}  Here, we calculate the 
mobility in the
presence of randomly distributed Coulomb impurity charges near the
surface with the electron-impurity interaction being screened by the
2D electron gas in the random phase approximation (RPA). The screened
Coulomb scattering is the only important scattering mechanism in our
calculation.  We assume that the direction of current flow is in the
$\hat{x}$ direction and that the magnetic field is in the $\hat{z}$
direction, with the graphene layer being in the 2D $xy$ plane.  In 
the presence of two types of carriers (electrons and
holes) in a finite magnetic field $B$, the current density in the
$\hat{x}$ and $\hat{y}$ directions are given by

\begin{eqnarray}
J_x & = & [\sigma_{xx}^{(e)}+\sigma_{xx}^{(h)}]E_x +
[\sigma_{xy}^{(e)}+\sigma_{xy}^{(h)}]E_y, \nonumber \\
J_y & = & [\sigma_{yx}^{(e)}+\sigma_{yx}^{(h)}]E_x +
[\sigma_{yy}^{(e)}+\sigma_{yy}^{(h)}]E_y. 
\end{eqnarray}
 The longitudinal conductivities are given by
\begin{equation}
\label{eq:conduct}
\sigma_{xx}^{(c)} = \sigma_{yy}^{(c)} =
\frac{\sigma_0^{(c)}}{1 +\left (\sigma_0^{(c)}R_H^{(c)}B \right )^2},
\end{equation}
where the superscript $c = e,h$ denotes electron 
and hole carriers, and the Hall conductivities are given by
\begin{equation}
\label{eq:sigma2}
\sigma_{xy}^{(c)}=-\sigma_{yx}^{(c)}=-\frac{\left [\sigma_0^{(c)}
  \right ]^2 R_H^{(c)}B}
{1 + \left (\sigma_0^{(c)}R_H^{(c)}B \right )^2},
\end{equation}
Here the zero-field electrical conductivity for each carrier is
defined by $\sigma_0^{(c)} = ({e^2}/{h})(2E_F^{(c)}\tau/\hbar)\approx
20 ({e^2}/{h}) (n_{(c)}/n_i)$ (see
Refs.~\onlinecite{kn:ando2006,kn:nomura2007,
kn:cheianov2006,kn:hwang2006c,kn:adam2007a} for details), and the Hall
coefficient $R_H^{(c)} = 1/n_{(c)}e^{(c)}$, where $n_{(c)}= n_e, n_h$
and $e^{(c)} = \pm e$, are the density and charge of the carriers,
respectively.

Shown in Fig.~\ref{Fig:rho} is the Boltzmann magnetoresistivity theory
(Eqs.~\ref{eq:conduct},~\ref{eq:sigma2}) for the two carrier model
that is compared with the gate-voltage tuned experimental results
reported in Ref.~\onlinecite{kn:schedin2006a} without any chemical doping.
The remarkable agreement with experimental data is further highlighted
in the inset where the theoretical longitudinal conductivities are
compared.  There is no free parameter in the theory (although
experimental results have been scaled by $n_i$, determined from the
conductivity dependence at high density).  For comparison, dashed
lines in the main panel of Fig.~\ref{Fig:rho} show resistivities with
one kind of carrier.  Since in general, the conductivities
($\sigma_{xx}$ and $\sigma_{xy}$) are proportional to the carrier
density $n$, the calculated resistivities for the single component
model diverge as $n \rightarrow 0$.  However, in two component model,
the Hall conductivity is given by $\sigma_{xy} = \sigma_{xy}^e +
\sigma_{xy}^h$ and $\sigma_{xy}^i \propto R_H^i$. Since the Hall
coefficient $R_H^i$ has different signs due to the sign of charge, the
sum of the Hall conductivity becomes zero when $n_e = n_h$.  This
results in a finite longitudinal magnetoresistivity and a zero Hall
resistivity at $n=0$. Similar to the case of zero
magnetic field, for finite magnetic field, the high density mobility 
is the only reliable measure of the concentration of 
Coulomb scattering centers. For graphene on a SiO$_2$ substrate we find
\begin{eqnarray}
\mu_{H} \equiv && \frac{\rho_{xy}}{\rho_{xx} B} = 
\frac{\sigma_{xy}}{\sigma_{xx}B} \nonumber \\
\approx && \frac{48.37}{n_i}
\left( \frac{n_e - n_h}{n_e + n_h} \right)
\stackrel{n > n_0}{\longrightarrow} \frac{48.37}{n_i}, 
\end{eqnarray}
where impurity density $n_i$ is measured in units
of $10^{10}~{\rm cm}^{-2}$ and $\mu_{H}$
is measured in units of ${\rm m}^{2}/{\rm Vs}$.

\section{Decreased mobility regime}
\label{Sect:mobdec}

\begin{figure}
\bigskip
\epsfxsize=0.9\hsize
\hspace{0.0\hsize}
\epsffile{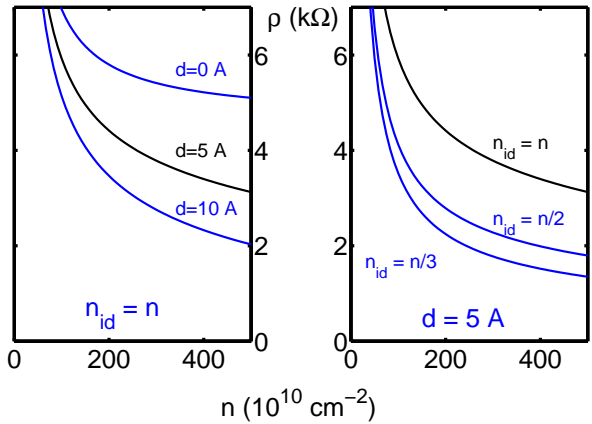}
\caption{\label{Fig:oldmodel}
(Color online) Left panel shows calculated density dependent resistivity for 
different locations of the impurity layer $d$ assuming that 
adsorbed chemicals contribute to the carrier density $n$ leaving
behind a charged impurity $n_{id}$ at a distance $d$ from the graphene
surface.  Right panel shows calculated resistivity using the
same model (with $d=5$ \AA) but assuming that only a fraction of 
the transferred charge contributes to the impurity density.}
\end{figure}

We first consider the most natural assumption for the effect of 
chemical adsorbents which is to transfer charge to the
graphene layer leaving behind a charged impurity at some distance $d
\sim 5$ \AA ~from the surface.  This is the 
``standard model'' assumed, for example, in carbon 
nanotubes.~\cite{kn:snow2006}  Shown in Fig.~\ref{Fig:oldmodel} are
the results of a Boltzmann calculation for this case.  
The calculation shown in the left
panel assumes ``conservation of charge'', i.e.  for every carrier
induced by the adsorbent there is left behind an impurity of equal 
charge that on the average is at a distance $d$ from the graphene
surface.  On the right panel we relax this assumption and for fixed
$d=5$ \AA, we show calculated results for the impurity charge $n_{id}$
being $n/2$ or $n/3$, where $n$ is the induced carrier density.
This model would always cause: (i) a decrease in mobility, because of
the increased impurity scattering, and (ii) an increase in carrier
density, because of the charge transferred by the adsorbed
molecule. However, both these conclusions are at odds with recent
experiments.~\cite{kn:schedin2006a}  This forces us to conclude that
the dominant source of scattering remains the random charged
impurities at the surface between graphene and the $SiO_2$ substrate,
and that the chemical adsorbent compensates some of this charged background
thereby providing an increased mobility.

\section{Increased mobility regime} 
\label{Sect:mobinc}

Within our magnetoresistance model, an increase in Hall mobility 
(for large carrier density)
necessarily implies a decrease in the charged impurity concentration,
which we argue is a consequence of adsorbed molecules acting as
compensators that partially neutralize the random Coulomb scatterers
in the substrate.  We note that molecular dopants may also shift the
threshold voltage as a consequence of surface chemistry effects and
this provides no information about the density of charged impurities
contributing to graphene transport.  An important goal of the current work 
is to theoretically explore 
the consequences of the compensation assumption using
the full Boltzmann transport theory and to make predictions for
subsequent field-effect conductivity measurements on the chemically
doped graphene sheets.  Since the prospect for applications of
graphene as sensors is an important technological possibility, our
work has practical implications.  More importantly, our work would
critically validate (or invalidate) the transport model and mechanism
in graphene

From the data reported in Ref.~\onlinecite{kn:schedin2006a}, one can
conclude that initial chemical doping serves to neutralize the charged
impurities in the substrate while not contributing to the carrier
density.  The experimental data motivates the assumption that there
is a roughly equal number of positive and negative charged impurities
so that $n_i^+ \sim n_i^-$.   This mechanism of adsorbed molecules
acting as compensators should eventually saturate, after which the
mobility would decrease as shown in Fig.~\ref{Fig:oldmodel}.  Before doping, 
the interface between graphene
and the substrate has a fixed density of random charged impurities
$n_i = n_i^+ + n_i^-$, comprising both positively ($n_i^+$) and
negatively ($n_i^-$) charged impurities.  The electron-hole asymmetry seen in
the experiments could be caused by the positive (negative) gate
voltage shifting positively (negatively) charged impurities by a
distance $d$ away from the surface.  We take this effect into account
explicitly, but unlike Ref.~\onlinecite{kn:hwang2006c}, we assume for
simplicity that before doping $n_i^+ = n_i^-$.  We also assume that 
$n_i^\pm = 0$ for graphene doped with $NH_3$ ($NO_2$).  This
model provides definitive predictions that could be tested in future
experiments.  For example, in the majority of samples where point
(i.e.  short-range) scattering is unimportant, chemical doping should
increase the electron-hole asymmetry with $NO_2$ ($NH_3$) showing
conductivity as a function of carrier density for electrons (holes)
that increases faster than the linear behavior predicted for holes
(electrons).  In the presence of both charge and point scattering,
applying a gate voltage results in a non-universal crossover between
linear and constant conductivity.  This correction caused by
short-range scatterers has been called the ``sub-linear''
conductivity, in contrast to the ``super-linear'' conductivity
described above to characterize the electron-hole asymmetry.  For 
samples with sub-linear behavior, chemical-doping would shift the
cross-over closer to the constant conductivity which would be seen
experimentally as shift of the onset of the sub-linear conductivity to
lower density.  If these features are seen in future experiments on
graphene, they would rule out alternative explanations including the
conventional wisdom in the nanotube community that chemical adsorbents
contribute directly to the graphene carrier density rather than
neutralize the substrate~\cite{kn:snow2006}, as we claim here.

\begin{figure}
\bigskip
\epsfxsize=0.9\hsize
\hspace{0.0\hsize}
\epsffile{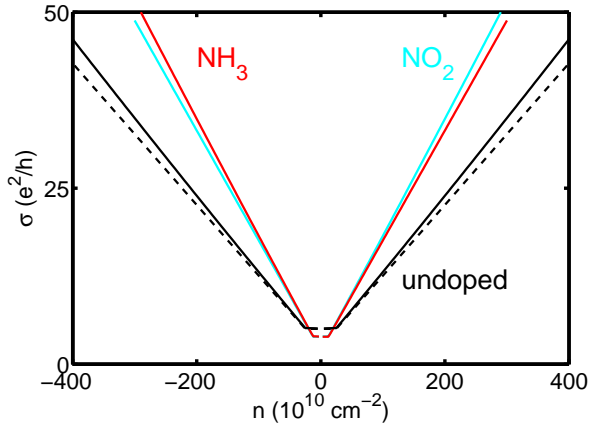}
\caption{\label{Fig:predict} (Color online) 
Shown (solid lines) are the calculated graphene 
conductivity for both doped and undoped cases for 
$n_i = 1.75 \times 10^{12}~{\rm cm}^{-2}$.  The asymmetry between 
$NH_3$ doping and $NO_2$  doping is caused by the voltage 
induced shifting of impurities as explained in the text.  
Also shown (dashed line) is the undoped case without
any voltage induced impurity shifting.  Note 
that for both $NH_3$ and $NO_2$, doping increases the sample mobility.}
\end{figure}

The predictions of our theoretical model are shown in 
Figs.~\ref{Fig:predict}.  First, we predict that 
because chemical doping neutralizes the
interface impurities, the high mobility should persist when doing
field-effect measurements after exposure to the adsorbant gases.  This
can be seen in Fig.~\ref{Fig:predict}, where the chemically-doped
cases have a higher slope than the undoped case.  Assuming that
point-scattering is still unimportant, the slope should continue to be
linear as in the undoped case.  Second, we predict a slight asymmetry
between hole doping ($NO_2$) and electron doping ($NH_3$) that is
caused by the same mechanism described above to explain the electron-hole
asymmetry in the undoped case.  The physical picture underlying our
transport model for graphene is simple: carrier transport in 2D
graphene layers is determined by charge impurity scattering with the
possibility of partially suppressing the impurity scattering through
adsorbate-induced compensation.  At low carrier density, both
electrons and holes are present, leading to the
observed~\cite{kn:schedin2006} magnetoresistive behavior.  If
short-range point scatterers are present in the system in addition to
charged Coulomb scatterers then the mobility will eventually
``saturate'' (i.e. become sublinear in carrier density) at high enough
carrier densities.

\section{Saturation of mobility}
\label{Sect:mobconst}

\begin{figure}
\bigskip
\epsfxsize=1\hsize
\hspace{0.0\hsize}
\epsffile{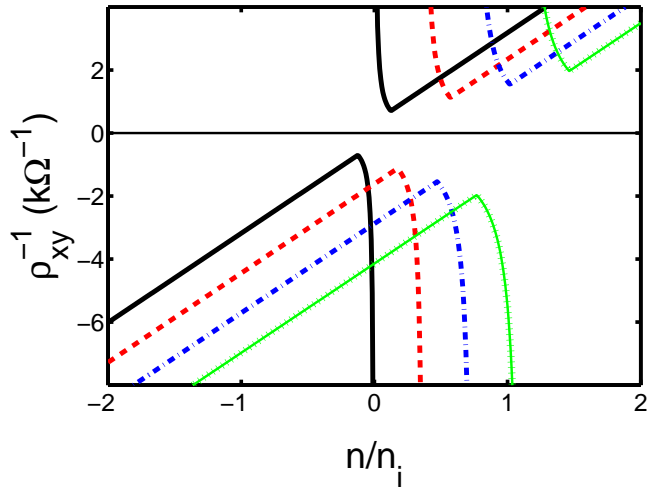}
\caption{\label{Fig:cond} (Color online) Main panel shows two carrier
model used to explain the results of 
Ref.~\protect{\onlinecite{kn:wehling2007}}.
Annealing causes a shift in the threshold voltage parameterized by
(from left to right) $\Delta/n_0=0,3,6,9$.}
\end{figure}

We observe that our two-carrier model also accounts phenomenologically
for the recent annealing experiments on molecular doped graphene
sheets reported in Ref.~\onlinecite{kn:wehling2007}.  In contrast to
Ref.~\onlinecite{kn:schedin2006a}, those experiments show no change in the
mobility implying that there is no change in the charged impurity
concentration $n_i$.  The experiments however do show a shift in
threshold voltage, which can be easily modeled by replacing $n$ in
Eq.~\ref{Eq:twochannel} with $\tilde n = n - (1 \mp \eta)\Delta$ for
electrons (holes).  Using $\eta \approx 0.2$ to characterize the
asymmetry in the data between electrons and holes, we find that this
reproduces all the features of the experiment including that for
electrons, the $1/\rho_{xy}$ traces are shifted in parallel to lower
values, that the peak corresponding to the divergence 
in $1/\rho_{xy}$ is also shifted to higher gate voltages, and is
broader and its minimum value increased.~\cite{kn:wehling2007}  These
features are shown in Fig.~\ref{Fig:cond}.  We point out that such features
involving different devices having different threshold voltages but
identical mobilities at high densities are quite common in Si MOSFETs and
GaAs 2D heterostructures.

\section{Conclusion}
\label{Sect:conclusion}

We have generalized the charged impurity scattering 
model (Ref.~\onlinecite{kn:hwang2006c}) of graphene transport
to calculate the low-field magnetoresistance and the chemical
doping situation as explored in recent experiments.~\cite{kn:schedin2006a,
kn:schedin2006}  We obtain reasonable agreement with experiments
by assuming that the chemical dopants lead to two distinct effects: 
increasing carrier density and decreasing (i.e. compensating)
the already existing charged impurity density.  The mobility
is a function of three parameters in the system: carrier 
density ($n$), impurity density ($n_i$), and impurity 
distribution (which can be parameterized in a simple model by 
a single length scale $d$ giving the separation of the charged
impurities in the substrate from the graphene carriers.  In our model, 
the effect of chemical doping on mobility must enter through the
modification of these three lengthscales: $d, n^{-1/2}, n_i^{-1/2}$.
How chemical doping affects these three parameters microscopically
is not known for the experiments of Refs.~\onlinecite{kn:schedin2006a,
kn:schedin2006}.  We have therefore made the simplest possible
assumption of keeping the unknown separation parameter $d$ 
a constant, assuming that chemical doping can only affect $n$ 
and $n_i$.  In principle, it is possible for chemical doping to
modify the impurity separation $d$, giving us another 
handle on the modification of mobility by chemical doping.  The 
important point to emphasize here is that to the extent 
chemical doping increases (decreases) the charged impurity
density $n_i$, it must change mobility $\mu \sim n_i^{-1}$.  If
$\mu$ remains unaffected as claimed in Refs.~\onlinecite{kn:schedin2006,
kn:wehling2007}, then chemical doping must somehow affect only 
$n$ without affecting $d$ and $n_i$ (the other possibility is 
a chemical doping induced complex interplay among $n$, $n_i$, $d$ 
and perhaps even the background dielectric constant $\kappa$, conspiring
to keep mobility a constant, a situation clearly beyond the scope of
any minimal theory).  We believe that the theory
presented herein provides the most straightforward explanation
of the experimental observations within the minimal model (and 
keeping the number of adjustable parameters to a minimum).  It
is gratifying in this context to point out that a recent
experimental study~\cite{kn:chen2007b} of systematic
potassium doping effects on graphene mobility 
in ultra-high vacuum obtains excellent agreement with our
minimal theory~\cite{kn:adam2007a,kn:hwang2006c} using
only $n_i$ as the single unknown parameter. 
   
In summary, we have developed a detailed microscopic theory for
graphene carrier transport which explains very well the existing
magnetoresistance data as well as accounting qualitatively and
quantitatively for the recently observed adsorbate-induced
modification of graphene transport properties which has implications
for application as sensors.  We make a number of specific experimental
predictions based on our transport model, whose validation (or
falsification) should further consolidate our understanding of 2D
carrier transport in graphene.  Our theory provides quantitative estimates
for the constraints on using graphene transport properties as chemical
sensors or detectors.  

We would like to thank A. Geim for sharing with us his unpublished
data and T. Einstein, M. Fuhrer and E. Williams for discussions. This
work was supported by U.S. ONR. {\em Note Added:} After submission of
this work~\cite{kn:hwang2006d}, alternate
mechanisms~\cite{kn:schedin2006,kn:titov2007,kn:wehling2007} for
graphene transport with chemical dopants have been proposed.   


\end{document}